\documentclass[12pt]{iopart}
\pdfoutput=1
\usepackage{graphicx,float}
\usepackage{xcolor}
\usepackage{iopams}
\begin{document}
\title[Effective dynamics in an asymmetric death-branching process]{Effective dynamics in an asymmetric death-branching process }
\author{Pegah Torkaman and Farhad H. Jafarpour}
\eads{\mailto{p.torkaman@basu.ac.ir}, \mailto{farhad@ipm.ir}}
\address{Physics Department, Bu-Ali Sina University, 65174-4161 Hamedan, Iran}
\vspace{10pt}
\begin{abstract}
In this paper we study activity fluctuations in an asymmetric death-branching process in one-dimension. 
The model, which is a variant of the asymmetric Glauber model, has already been studied in~\cite{TJ15a}. It is known that in the low-activity region i.e. below the typical activity 
in the steady-state, the dynamical free energy of the system can be calculated exactly. However, the behavior of the system 
in the high-activity region is different and more interesting. The system undergoes a series of dynamical phase transitions. 
In present work we justify the hierarchy of dynamical phase transitions in terms of effective interactions in the system. 
It turns out that the effective interactions are long-range and that they can be described in terms of interactions between repelling shock fronts. 
\end{abstract}
\vspace{2pc}
\maketitle
\section{Introduction}
\label{I}

Recent studies have highlighted the importance of rare events in different fields of science. These rare events or large deviations from the typical 
behavior, despite having very small probability of occurrence, might cause extreme consequences such as climate extremes, earthquakes and other potentially catastrophic phenomena~\cite{rare1}. Some classical examples include 
phase transformation and protein-folding~\cite{rare2}. 

Markov processes are used to model a variety of important random systems. Recently much attention has been devoted to conditioning a Markov 
process on a rare event. Conditioning a Markov process on a rare event means fixing the average of a given dynamical observable on an atypical 
value. On the other hand, there is a conditioning-free process for which the average of a given dynamical observable in large time limit is equal to that 
of the conditioned process. This conditioning-free process is sometimes called the effective or driven process~\cite{CT15}. 

Apart from the dynamics of classical and quantum stochastic systems, their dynamical phase behavior can be studied using 
the thermodynamics of trajectories, sometimes known as Ruelle's thermodynamics~\cite{Ruelle}. The fluctuations of the dynamical observable
or equivalently a time-extensive order parameter, within an ensemble of trajectories are described by large-deviation rate functions which 
play the role of dynamical free-energies. The dynamical phase transitions are controlled by a biasing field which is coupled to the order parameter. 
Instead of fixing the average of the dynamical observable in a conditioned Markov process one can fix the value of its corresponding biasing field~\cite{LAV07}. 

Generally speaking, the ensemble average of a given dynamical observable in a long observation period, might depend on time. So the time-translation 
invariance might be broken.  However, there is a time interval, far from the initial and final observation time, in which the ensemble average 
of the dynamical observable is independent of time and the time-translation invariance holds. The biased trajectories of the original process under 
investigation in this time-translational invariant regime coincide with unbiased trajectories of the effective process. The rare trajectories of interest in 
the original process can be characterized as typical trajectories for the effective process. The effective process consists of new set of interactions
called effective interactions, which has to be imposed on the original process to make atypical behavior typical~\cite{JS10}.

 It is definitely of great interest to understand the dynamics of the systems when a rare event takes place. However, finding analytical expressions
  for the effective interactions is generally quite difficult, although some limited efforts have recently been done. There are only a couple of examples for which the effective interactions have been calculated exactly~\cite{JS10,PS11,JS14,HMS15}. We are involved with complex systems 
which have many degrees of freedom. Therefore even if the exact result is obtained, such a large dimensionality might not give direct information 
about the physical nature of effective interactions in the system; however, it might shed some light to understanding of the physics of the problem.
An interesting aspect about the nature of the effective interactions is that they are usually very complicated and non-local~\cite{JS15}. 
In a recent paper~\cite{TJ15b} the 
authors have shown that under some conditions the effective interactions in the effective process can be local as the original process.

In a recent paper~\cite{TJ15a} the authors have studied the fluctuations of activity, defined as the number of configuration changes of the system in a dynamical trajectory, and the dynamical phase transitions in a stochastic system of classical particles consisting of asymmetric branching and death processes connecting to two particle reservoirs at the boundaries of an one-dimensional lattice. It has been shown that the system indeed undergoes both continuous and discontinuous dynamical phase transitions for an atypical value of activity below its typical value in the steady-state. The dynamics of the system conditioned on a lower-than-typical activity region has been investigated and different phases have been characterized according to the configuration of the system at the beginning and the end of each trajectory during the observation time. 

In this paper, considering the same system and the same dynamical observable, we study the effective interactions which generate the atypical values of activity. We would also investigate how these effective interactions lead to a new dynamical phase behavior especially in the higher-than-typical activity region where a hierarchy of dynamical phase transitions occur. It turns out that effective dynamics for any value of the biasing field conjugated to the dynamical observable is governed by the interactions between shock fronts which perform simple random walk on the lattice. However, the number of shock fronts which are involved depends on the value of the biasing field. In other words, as the average activity increases the number of shock fronts involved in the effective dynamics increases. This can be easily understood from the fact that, as we will see, the activity of the system is merely generated by the shock fronts. 

\section{Basic concepts and the model}
\label{II}
We start with a continuous-time Markov process with a finite configuration space $ \{C\}$ in which a spontaneous transition from configuration $C$ to $C'$ takes place with a transition rate $\omega_{C \to C'}$. Considering a complete basis vector $\{ | C \rangle \}$, the probability 
$P(C,t)=\langle C | P(t) \rangle$ of finding the system in configuration $C$ at time $t$ satisfies the following master equation~\cite{S01}
\begin{equation}
\label{ME}
\frac{d}{dt} | P(t) \rangle = \hat{\cal{H}} | P(t)\rangle
\end{equation}
where $\hat{\cal{H}} $ is the stochastic generator of the Markov process with matrix elements
$$
\langle C | \hat{\cal{H}} | C' \rangle = \omega_{C' \to C} - \delta_{C,C'} \sum_{C'' \neq C} \omega_{C \to C''} 
$$
in which we have assumed $\omega_{C \to C}=0$. In order to study the dynamics of the system on a value 
of a given dynamical observable, which is sometimes referred to as conditioning a Markov process on a rare event which is characterized 
by an atypical value of the observable, we start with the following moment generating function $\langle e^{-s K} \rangle =\sum_{K} P(K) e^{-s K} $ 
for the activity $K$ of system which can be written as
\begin{equation}
\langle e^{-s K} \rangle = \langle 1 | P_s(t) \rangle
\end{equation}
in which $ \langle 1 |=\sum_{C} \langle C |$ is a summation vector and $ | P_s(t) \rangle$ satisfies the following master equation
\begin{equation}
\label{SME}
\frac{d}{dt} | P_s(t) \rangle = \hat{\cal{H}}(s) | P_s(t)\rangle 
\end{equation}
The activity is a time-integrated current which is defined as 
the number of configuration changes during the finite time interval $[0,t]$. The operator $\hat{\cal{H}}(s) $ is a non-stochastic generator, known 
as the modified or tilted Hamiltonian, which can be constructed by multiplying all of the non-diagonal elements of $\hat{\cal{H}}$ by a factor $e^{-s}$ as follows~\cite{LAV07,JS10}
$$
\langle C | \hat{\cal{H}}(s) | C' \rangle =e^{-s} \omega_{C' \to C} - \delta_{C,C'} \sum_{C'' \neq C} \omega_{C \to C''} \; 
$$
The parameter $s$ is a counting filed conjugated to the activity $K$. Fixing $s$ on some non-zero values corresponds to 
studying the dynamics of the system conditioned on some atypical value of activity $K$. In fact, it plays the role of a biasing field in 
the ensemble of dynamical trajectories, known as s-ensemble~\cite{HJGC09}. The sum of unnormalized probabilities $P(C,s,t)=\langle C | P_s(t) \rangle$ 
gives the dynamical partition function of the s-ensemble defined as $Z(s,t)= \langle 1 |P_s(t) \rangle$. The logarithm of this partition function is called the dynamical free energy of the system whose singularities determine dynamical phase behavior of the system under investigation~\cite{LAV07}.

Let us consider the following eigenvalue equations for $\hat{\cal{H}}(s)$
\begin{equation}
\label{EVE}
\begin{array}{l}
\hat{\cal{H}}(s) | \Lambda(s) \rangle = \Lambda (s) | \Lambda(s) \rangle \; , \\
\langle \tilde{\Lambda}(s) | \hat{\cal{H}}(s) = \Lambda (s) \langle \tilde{\Lambda}(s) | \; .
\end{array}
\end{equation}
We denote the largest eigenvalue of $\hat{\cal{H}}(s)$ by  $\Lambda^{{\ast}}(s)$ and the left and right eigenvectors 
corresponding to this eigenvalue by  $\langle \tilde{\Lambda}^{{\ast}}(s) |$ and  $| \Lambda^{{\ast}}(s) \rangle $ respectively. 
It can be shown that in the large time limit and for a system with finite configuration space the generating function $\langle e^{-s K} \rangle$ 
(or the dynamical partition function $Z(s,t)$) has a large deviation form of the type $\langle e^{-s K} \rangle \asymp e^{t \Lambda^{{\ast}}(s)}$ in which
the sign $\asymp$ is interpreted as expressing an equality relationship on a logarithmic scale~\cite{T09}. 
This means that $\Lambda^{{\ast}}(s)$ is the dynamical free energy of system whose derivatives with respect to $s$ determine the dynamical 
phase structure.
  
As it was explained in the introduction, there is a conditioning-free representation for a conditioned Markov process, known as effective or driven process. 
The generator of this stochastic process can be obtained from the non-stochastic generator $\hat{\cal{H}}(s)$ using a
generalization of Doob's h-transformation~\cite{CT15} 
\begin{equation}
\label{relation}
\hat{\cal{H}}_{eff}(s) =\hat{U} \hat{\cal{H}}(s) \hat{U}^{-1}-\Lambda^{{\ast}}(s) 
\end{equation}
where operator $\hat{U}$ is a diagonal matrix with the matrix element
$ 
\langle C | \hat{U} | C \rangle =  \langle \tilde{\Lambda}^{{\ast}}(s) |C \rangle \; . 
$
The matrix elements of the effective Hamiltonian $\hat{\cal{H}}_{eff}(s)$ are given by
\begin{equation}
\label{element}
\langle C | \hat{\cal{H}}_{eff}(s) | C' \rangle =\frac{
 \langle \tilde{\Lambda}^{{\ast}}(s) |C \rangle
\langle C | \hat{\cal{H}}(s) | C' \rangle}
{ \langle \tilde{\Lambda}^{{\ast}}(s) |C' \rangle} 
-\delta_{C,C'} \Lambda^{{\ast}}(s)  \; .
\end{equation}

In what follows we will define a model which was introduced and studied first in~\cite{MTJ14}. Let us consider a system of classical particles 
defined on a one-dimensional lattice of length $L$ with open boundaries. The stochastic evolution of the system in time is determined 
by the following reaction rules in the bulk of the lattice
\begin{equation}
\label{rules}
\begin{array}{ll}
A \; \emptyset \; \longrightarrow \; \emptyset \; \emptyset \quad \mbox{with the rate} \quad \omega_1 \\
A \; \emptyset \; \longrightarrow \; A \; A \quad \mbox{with the rate} \quad \omega_2.
\end{array}
\end{equation}
in which a particle (vacancy) is labeled with $A$ ($\emptyset$).
The reaction rules at the boundaries are also given by
\begin{equation}
\label{rules2}
\begin{array}{ll}
\emptyset \; \longrightarrow \; A \quad \mbox{with the rate} \quad \alpha \quad \mbox{at the left boundary} \quad \\
A \;  \longrightarrow \; \emptyset \quad \mbox{with the rate} \quad \beta \quad \mbox{at the right boundary} \; .
\end{array}
\end{equation}

This model can be regarded as an asymmetric variant of the zero-temperature Glauber model on a one-dimensional lattice with open boundaries in which 
the transition rates of two reaction rules defined as
$\emptyset \;A \; \longrightarrow \; \emptyset \; \emptyset$
and
$ \emptyset \; A \; \longrightarrow \; A \; A$
are equal to zero \cite{KhA01,KJSh}.
The stochastic generator of the system is given by
\begin{equation}
\label{hamiltonian1}
\hat{\cal{H}} =  \hat{\cal{L}}  \otimes {\cal I}^{\otimes (L -1)}  + \sum_{i=1}^{L-1} {\cal I}^{\otimes (i-1)}\otimes \hat{h} \otimes {\cal I}^{\otimes (L-i-1)}
    + {\cal I}^{\otimes (L -1)}\otimes \hat{\cal{R}}
\end{equation}
in which $\cal I$ is a $2 \times 2$ identity matrix. Introducing the basis kets
$$
\vert  \emptyset \rangle = \left( \begin{array}{c}
1\\
0\\
\end{array} \right)\, ,\;\;  
\vert A \rangle=\left( \begin{array}{c}
0\\
1\\
\end{array} \right)\, \; ,
$$
the matrix $\hat{h}$ in the basis of $\{\emptyset \emptyset,\emptyset A,A \emptyset,AA\}$ and the matrices $\hat{\cal{L}}$ and $\hat{\cal{R}}$ 
in the basis of $\{\emptyset,A\}$ can be written as follows
$$
\hat{h}=\left( \begin{array}{cccc}
0 & 0&\omega_{1}&0\\
0 & 0&0&0\\
0&0&-(\omega_{1}+\omega_{2})&0\\
0&0&\omega_{2}&0\\
\end{array} \right)\, , 
\hat{\cal L} =\left( \begin{array}{cc}
-\alpha & 0\\
\alpha & 0\\
\end{array} \right)\, , 
\hat{\cal R}=\left( \begin{array}{cc}
0 & \beta\\
0 & -\beta\\
\end{array} \right)\, . 
$$
Considering the activity as the proper dynamical observable, the modified hamiltonian of the system can easily be written as
\begin{equation}
\label{hamiltonian2}
\hat{\cal{H}}(s)  =  \hat{\cal{L}}(s)  \otimes {\cal I}^{\otimes (L -1)}+\sum_{i=1}^{L-1} {\cal I}^{\otimes (i-1)}\otimes \hat{h}(s) \otimes {\cal I}^{\otimes (L-i-1)}
      + {\cal I}^{\otimes (L -1)}\otimes \hat{\cal{R}}(s) 
\end{equation}
where
$$
\hat{h}(s)=\left( \begin{array}{cccc}
0 & 0&\omega_{1}e^{-s}&0\\
0 & 0&0&0\\
0&0&-(\omega_{1}+\omega_{2})&0\\
0&0&\omega_{2}e^{-s}&0\\
\end{array} \right)\, ,
\hat{\cal{L}}(s) =\left( \begin{array}{cc}
-\alpha & 0\\
\alpha e^{-s} & 0\\
\end{array} \right)\, , 
\hat{\cal{R}}(s)=\left( \begin{array}{cc}
0 & \beta e^{-s}\\
0 & -\beta\\
\end{array} \right)\, . 
$$

As we explained above, it is known that fixing the counting field $s$ corresponds to the study of an atypical value of the activity. 
The typical value of the activity in the steady-state of this model has already been calculated analytically in~\cite{TJ15a}. In the following sections 
we will first solve the eigenvalue problem~(\ref{EVE}). It turns out that the full spectrum of the modified Hamiltonian~(\ref{hamiltonian2}) 
can be calculated exactly; therefore, the largest eigenvalue is recognized and then the left and right eigenvectors corresponding 
to that eigenvalue are calculated analytically. Finally exact analytical results for the effective ineractions of the system will be 
obtained and their physical interpretation will also be discussed.   

\section{Diagonalization of $\hat{\cal{H}}(s)$}
\label{III}
\begin{figure}[!t]
\centering
\includegraphics[scale=1]{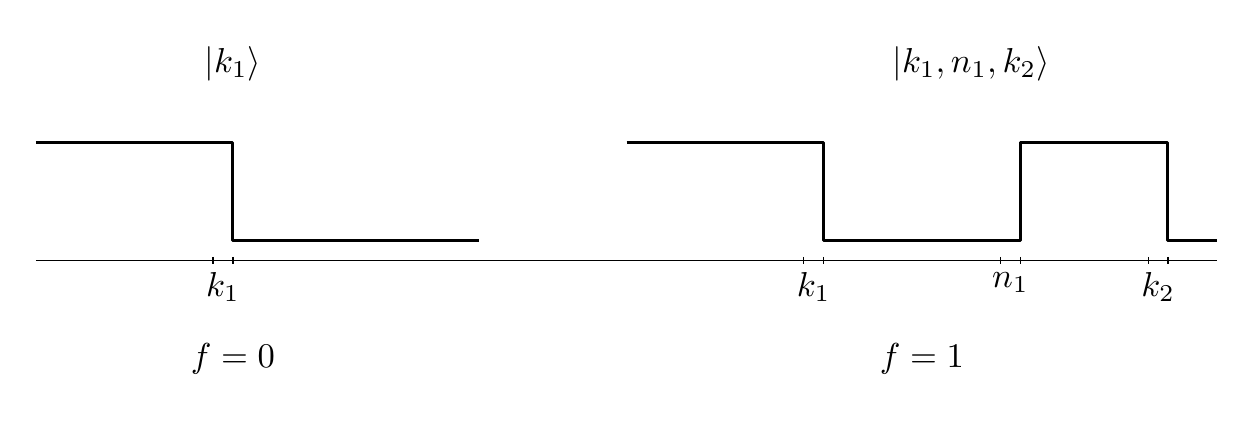}
\caption{Simple sketch of the base vectors defined in (\ref{1shf}) and (\ref{3shf}).}
\label{fig1}
\end{figure}

In this section we will show that the modified Hamiltonian~(\ref{hamiltonian2}) can be diagonalized exactly. Without loss of generality we 
assume that the system size $L$ is an even number. Let us consider a new complete basis vector 
\begin{equation}
\label{newbasis0}
\{ { \vert k_1 \rangle},{ \vert k_1,n_1,k_2 \rangle},\cdots, { \vert k_1,n_1,k_2,n_2,\cdots,n_{\frac{L}{2}},k_{\frac{L}{2}+1} \rangle}\}
\end{equation}
with $2^L$ members in which 
\begin{eqnarray}
\label{newbasis}
\vert k_{1},n_{1},k_{2},n_{2},\cdots,n_{f},k_{f+1} \rangle &=&
\vert A \rangle^{\otimes k_{1}} 
\otimes 
\vert \emptyset \rangle^{\otimes (n_{1}-k_{1})}
\otimes 
\vert A \rangle^{\otimes (k_{2}-n_{1})}
\otimes 
\cdots   \nonumber \\
&\otimes &
\vert A \rangle^{\otimes (k_{f+1}-n_{f})}
\otimes 
\vert \emptyset \rangle^{\otimes (L-k_{f+1})} 
\end{eqnarray}
and that $0 \le k_{1} < n_{1} < k_{2} < \cdots<n_{f} < k_{f+1} \le L$. Each member of~(\ref{newbasis}) can be thought of as a product shock measure 
with $N=2f+1$ shock fronts. The dimensionality of the subspace spanned by the vectors $\{ \vert k_{1},n_{1},k_{2},n_{2},\cdots,n_{f},k_{f+1} \rangle \}$ 
is
$$
{\cal C}_{L+1,N} \equiv \frac{(L+1)!}{N!(L+1-N)!}\; .
$$
As an example, let us write~(\ref{newbasis}) for $f=0$ as
\begin{eqnarray} 
\label{1shf}
|k_1 \rangle = \vert A \rangle^{\otimes k_1} \otimes \vert \emptyset \rangle^{\otimes (L-k_1)} \; 
\end{eqnarray}
where $0 \le k_1 \le L$ and for $f=1$ as
\begin{eqnarray} 
\label{3shf}
 |k_1,n_1,k_2 \rangle = \vert A \rangle^{\otimes k_1} \otimes \vert \emptyset \rangle^{\otimes (n_1-k_1)}\otimes \vert A \rangle^{\otimes (k_2-n_1)} 
\otimes \vert \emptyset \rangle^{\otimes (L-k_2)} 
\end{eqnarray}
where $0 \le k_1 < n_1 < k_2 \le L$. In~FIG.\ref{fig1} 
we have plotted $|k_1 \rangle$ with $f=0$ and $|k_1,n_1,k_2 \rangle$ with $f=1$ schematically. 
From the dynamical rules~(\ref{rules}) it is clear that under the time evolution generated by~(\ref{hamiltonian1}) only the the shock 
fronts at the positions denoted by $k_i$'s move while the shock fronts at the positions denoted by $n_i$'s do not move. This is the case for the generator 
(\ref{hamiltonian2}) as well. The structure of $\hat{\cal{H}}(s)$ in the new basis is as follows
\begin{equation}
\label{new modified hamiltonian}
\hat{\cal{H}}(s)=
\left( \begin{array}{ccccccc}
A_{1} & B_{3}&0&0&\cdots&0&0\\
0 & A_{3}&B_{5}&0&\cdots&0&0\\
0&0&A_{5}&B_{7}&\cdots&0&0\\
\vdots&&&\ddots&\ddots&& \vdots\\
0&0&0&0&\cdots&B_{L-1}&0\\
0&0&0&0&\cdots&A_{L-1}&B_{L+1}\\
0&0&0&0&\cdots&0&A_{L+1}
\end{array} \right)\,
\end{equation}
where the dimensionality of the matrix $A_N$ is ${\cal C}_{L+1,N} \times {\cal C}_{L+1,N}$
and that of $B_N$ is ${\cal C}_{L+1,N-2} \times {\cal C}_{L+1,N}$ with $N=1,3,5,\cdots,L+1$.
It is easy to check that  
$$
 \sum_{N=1 , N \in odd}^{L+1}{\cal C}_{L+1,N} =2^L 
$$
which confirms the dimensionality of the total state space is counted correctly. 

The structure of the modified Hamiltonian (\ref{new modified hamiltonian})
depicts the picture in which there are $\frac{L}{2}+1$ invariant sectors. 
Each sector corresponds to a group of product shock measures which contains up to
a certain number of shock fronts. For example acting 
 (\ref{new modified hamiltonian}) on $|k_1,n_1,k_2 \rangle$ with 
 $0 \le k_1 < n_1 < k_2 \le L$, which is a product shock measure with three shock fronts, 
 gives a linear combination of the product shock measures with a single or three shock fronts;
 therefore, the number of shock fronts on the lattice might remain unchanged or be reduced
  by the evolution generator $\hat{\cal{H}}(s)$. Note that their number can never increase.

It has also been found that acting the modified Hamiltonian (\ref{new modified hamiltonian})
on each of the product shock measures (\ref{newbasis}) with $N=2f+1$ shock fronts,
gives a series of evolution equations which are quite similar to those of $f+1$ biased
random walkers at the lattice sites $\{k_1,k_2,\dots,k_{f+1}\}$ besides $f$
obstacles at the lattice sites $\{n_1,n_2,\dots,n_f\}$ given that the random walkers
are at least one lattice site away from the obstacles. Each pair of consecutive random walker and obstacle can disappear simultaneously whenever they collide with each other. 
Each random walker can also reflect from an obstacle, as the first and the last 
random walkers reflect from the boundaries of the lattice.  

As a matter of fact, we are dealing with a counting process \cite{HSh07} in which measuring of the activity, as the number of configuration changes or microscopic transitions of the system in a dynamical trajectory, is requested; therefore, the counting field $s$ is introduced in (\ref{hamiltonian2}). By studying the structure of the modified Hamiltonian 
(\ref{new modified hamiltonian}) in the new basis (\ref{newbasis0}),
we have found that multiplying all non-diagonal elements of (\ref{hamiltonian1}) by an exponential factor $e^{-s}$, corresponds to tilting of the jump rates of the moving shock fronts (random walkers) to the left or right and annihilation rate of them with the obstacles by the same factor $e^{-s}$. 
This means that the activity of the system is generated by 
the moving shock fronts at the lattice sites $\{k_1,k_2,\dots,k_{f+1}\}$.

Now, in order to diagonalize of the modified Hamiltonian (\ref{new modified hamiltonian}),
we start with diagonalizing $A_N$ for an arbitrary 
$N$. Since $A_N$ is not symmetric we have
\begin{eqnarray}
\label{EVP}
A_N | a_{i_N}^{(N)} \rangle &=& a_{i_N}^{(N)} | a_{i_N}^{(N)} \rangle \; , \nonumber \\ \\ 
\langle \tilde{a}_{i_N}^{(N)} | A_N &=& a_{i_N}^{(N)} \langle \tilde{a}_{i_N}^{(N)}| \nonumber
\end{eqnarray}
where $i_N=1,2,\cdots,{\cal C}_{L+1,N}$. Solving these eigenvalue equations determines the full spectrum of the modified Hamiltonian
\begin{eqnarray}
\label{ev}
\Lambda (s) =\{ \{a_{i_1}^{(1)}\},\{a_{i_3}^{(3)}\},\cdots,\{a_{i_{L+1}}^{(L+1)}\} \} \; .
\end{eqnarray}
On the other hand, solving~(\ref{EVP}) results in finding the exact expressions for the left and right eigenvectors of $\hat{\cal H}(s)$
so that for $\Lambda (s)=a_{i_N}^{(N)}$ we have
\begin{eqnarray}
\label{eigen2}
| \Lambda(s) \rangle & =& \bigoplus _{ l=1 , l \in odd}^{L+1} |{ \cal{X} }_{i_N,l}^{(N)} \rangle \; , \nonumber \\ \\
\langle \tilde{\Lambda}(s) | &=& \bigoplus _{l=1 , l \in odd}^{L+1} \langle{ \tilde{\cal{X}} }_{i_N,l}^{(N)} \vert \nonumber
\end{eqnarray}
in which
\begin{eqnarray}
\label{eigen3}
|{ \cal{X} }_{i_N,l}^{(N)} \rangle =\left\{
\begin{array}{ll}
0                                              & \quad \mbox{for}\quad l > N \; , \\   \\
| a_{i_N}^{(N)} \rangle             & \quad \mbox{for} \quad l=N\; , \\ \\
\Pi _{j=l,j \in odd}^{N-2} (a_{i_N}^{(N)} -A_j)^{-1} B_{j+2} | a_{i_N}^{(N)} \rangle
                                               & \quad \mbox{for} \quad l < N \; ,
\end{array}
\right.
\end{eqnarray}
and
\begin{eqnarray}
\label{eigen4}
\langle{ \tilde{\cal{X}} }_{i_N,l}^{(N)} \vert =\left\{
\begin{array}{ll}
\langle \tilde{a}_{i_N}^{(N)}|  \Pi _{j=N+2,j \in odd}^{l} B_{j} (a_{i_N}^{(N)} -A_j)^{-1} 
                                                         & \quad \mbox{for}\quad l > N \; , \\   \\
\langle \tilde{a}_{i_N}^{(N)}|              & \quad \mbox{for} \quad l=N\; , \\ \\
0                                                       & \quad \mbox{for} \quad l < N \; .
\end{array}
\right. 
\end{eqnarray}
It turns out that for any arbitrary $N$, $A_N$ can be diagonalized using the standard plane wave ansatz. However, before diagonalizing the most
general case let us first solve the two simplest sectors $f=0$ and $f=1$. 
For $f=0$ one starts with
\begin{eqnarray}
\label{f=0}
\vert a_{i_1}^{(1)} \rangle & = & \sum_{k_{1}=0}^L C^{i_1}_{k_{1}} \vert k_{1} \rangle  \; , \nonumber \\ 
\langle \tilde{a}_{i_1}^{(1)} \vert & = & \sum_{k_{1}=0}^L \tilde{C}^{i_1}_{k_{1}} \langle k_{1} \vert 
\end{eqnarray}
for $i_1=1,2,\cdots,{\cal C}_{L+1,1}$. Defining 
$$
\eta \equiv \sqrt{\frac{\omega_{2}}{\omega_{1}}},
\quad \zeta \equiv 1-\frac{\alpha}{\omega_{2}},
\quad \xi \equiv 1-\frac{\beta}{\omega_{1}}
$$
and
$$
F(x,y,z) \equiv  (x+x^{-1}z)e^{-s}-(y z+y^{-1}) 
$$
and considering a simple plane wave ansatz for the coefficients $C^{i_1}_{k_{1}}$ and $\tilde{C}^{i_1}_{k_{1}}$ the eigenvalues of $A_1$ are obtained to be
\begin{equation}
\label{eigen f=0}
a_{i_1}^{(1)}=-(\omega_{1}+\omega_{2})+e^{-s}\sqrt{\omega_{1} \omega_{2}}(z_{1}+z_{1}^{-1}).
\end{equation}
and that the coefficients in~(\ref{f=0}) are given by
\begin{eqnarray}
&& C^{i_1}_{k_{1}}=\frac{\eta^{k_{1}}}{(1-\zeta)^{\delta_{k_{1},0}}(1-\xi)^{\delta_{k_{1},L}}} \Big(\psi(z_{1}) z_{1}^{k_{1}}+\psi(z_{1}^{-1})z_{1}^{-k_{1}}\Big) \; , \nonumber \\
&& \tilde{C}^{i_1}_{k_{1}}=\eta^{-k_{1}} \Big(\psi(z_{1}) z_{1}^{k_{1}}+\psi(z_{1}^{-1})z_{1}^{-k_{1}}\Big) \; . \nonumber 
\end{eqnarray}
One also finds that the following constraints have to be satisfied
$$
\frac{\psi(z_{1})}{\psi(z_{1}^{-1})}=-\frac{F(z_{1},\eta,\zeta)}{F(z_{1}^{-1},\eta,\zeta)} =-z_{1}^{-2L}\frac{F(z_{1}^{-1},\eta^{-1},\xi)}{F(z_{1},\eta^{-1},\xi)}  
$$
which give an equation for $z_1$ as follow
$$
z_{1}^{2L}=\frac{F(z_{1}^{-1},\eta,\zeta)F(z_{1}^{-1},\eta^{-1},\xi)}{F(z_{1},\eta,\zeta)F(z_{1},\eta^{-1},\xi)} \; .
$$
The eigenvalues of $A_1$ or $f=0$ sector can be obtained by inserting the solutions of this equation for $z_1$ in~(\ref{eigen f=0}).
The same procedure can be used for diagonalizing $A_3$ or $f=1$ sector. The right and left eigenvectors of $A_3$ can be written as
\begin{eqnarray}
\label{f=1}
&& \vert a_{i_3}^{(3)} \rangle=\sum_{k_{1}=0}^{n_{1}-1}\sum_{k_{2}=n_{1}+1}^{L} 
C^{i_3}_{k_{1},n_{1},k_{2}} \vert k_{1},n_{1},k_{2} \rangle  \; , \nonumber \\
&& \langle \tilde{a}_{i_3}^{(3)} \vert=\sum_{k_{1}=0}^{n_{1}-1}\sum_{k_{2}=n_{1}+1}^{L} 
 \tilde{C}^{i_3}_{k_{1},n_{1},k_{2}} \langle k_{1},n_{1},k_{2} \vert
\end{eqnarray}
for every $n_1$ which satisfies $0 \le k_1 < n_1 < k_2 \le L$ and $i_3=1,2,\cdots,{\cal C}_{L+1,3}$. The eigenvalues of $A_3$ is given by
\begin{equation}
\label{eigen f=1}
a_{i_3}^{(3)}=-2(\omega_{1}+\omega_{2})+e^{-s}\sqrt{\omega_{1} \omega_{2}}(z_{1}+z_{2}+z_{1}^{-1}+z_{2}^{-1}).
\end{equation}
and the coefficients in~(\ref{f=1}) are obtained to be
\begin{eqnarray}
&& C^{i_3}_{k_{1},n_{1},k_{2}}=\prod_{i=1}^{2} \frac{\eta^{k_{i}}}{(1-\zeta)^{\delta_{k_{1},0}}(1-\xi)^{\delta_{k_{2},L}}}
\Big( \psi_i(z_i)z_i^{k_{i}}+\psi_i(z_i^{-1}) z_i^{-k_{i}} \Big ) \; , \nonumber \\ 
&& \tilde{C}^{i_3}_{k_{1},n_{1},k_{2}}=\prod_{i=1}^{2} \eta^{-k_{i}} \Big( \psi_i(z_i)z_i^{k_{i}}+\psi_i(z_i^{-1}) z_i^{-k_{i}} \Big ) \; .\nonumber 
\end{eqnarray}
The constraints in this case are 
\begin{eqnarray}
&& \frac{\psi_1(z_1)}{\psi_1(z_1^{-1})}=-\frac{F(z_1,\eta,\zeta)}{F(z_1^{-1},\eta,\zeta)}=-z_1^{-2n_{1}} \; ,   \nonumber  \\ 
&& \frac{\psi_2(z_2)}{\psi_2(z_2^{-1})}=-z_2^{-2L}\frac{F(z_2^{-1},\eta^{-1},\xi)}{F(z_2,\eta^{-1},\xi)}=-z_2^{-2n_{1}}   \nonumber 
\end{eqnarray}
for every allowed $n_1$. Finally the equations for $z_1$ and $z_2$ for every given $n_1$ which lies in $0 \le k_1 < n_1 < k_2 \le L$ are
\begin{eqnarray}
&& z_1^{2 n_{1}}= \frac{F(z_1^{-1},\eta,\zeta)}{F(z_1,\eta,\zeta)} \; , \nonumber  \\
&& z_2^{2(L- n_{1})}= \frac{F(z_2^{-1},\eta^{-1},\xi)}{F(z_2,\eta^{-1},\xi)}  \; . \nonumber 
\end{eqnarray}
At the end of this section we bring the results of diagonalizing $A_N$. Quite similar to what we have already done for diagonalizing $A_1$ and $A_3$ 
one can generalize the procedure to find that the eigenvectors of $A_N$ can be written as
\begin{eqnarray}
\label{final eigenvectors}
&& \vert a_{i_N}^{(N)} \rangle=\sum_{k_{1}}\sum_{k_{2}}\cdots
\sum_{k_{f+1}} C^{i_N}_{k_{1},n_{1},k_{2},n_{2},\cdots,k_{f+1}} \vert k_{1},n_{1},k_{2},n_{2},\cdots,k_{f+1}\rangle  \; , \nonumber \\ 
&& \langle \tilde{a}_{i_N}^{(N)} \vert=\sum_{k_{1}}\sum_{k_{2}}\cdots
\sum_{k_{f+1}} \tilde{C}^{i_N}_{k_{1},n_{1},k_{2},n_{2},\cdots,k_{f+1}} \langle k_{1},n_{1},k_{2},n_{2},\cdots,k_{f+1} \vert
\end{eqnarray}
for every set of $\{n_1,\cdots ,n_{f}\}$ which satisfies $0 \le k_{1} < n_{1} < k_{2} < \cdots<n_{f} < k_{f+1} \le L$ and that $i_N=1,2,\cdots,{\cal C}_{L+1,N}$. 
Straightforward calculations give  the eigenvalues of $A_N$
\begin{equation}
\label{ev2}
a_{i_N}^{(N)}=-(f+1)(\omega_1+\omega_2)+e^{-s}\sqrt{\omega_1 \omega_2}\sum_{i=1}^{f+1}(z_i+z_i^{-1}) \; .
\end{equation}
The coefficients in~(\ref{final eigenvectors}) are also found to be
\begin{eqnarray}
\label{coef}
&& C^{i_N}_{k_{1},n_{1},k_{2},n_{2},\cdots,k_{f+1}}=\prod_{i=1}^{f+1} 
\frac{\eta^{k_{i}}}{(1-\zeta)^{\delta_{k_{1},0}}(1-\xi)^{\delta_{k_{f+1},L}}}\Big( \psi_i(z_i)z_i^{k_{i}}+\psi_i(z_i^{-1}) z_i^{-k_{i}} \Big )\; , \nonumber  \\
&& \tilde{C}^{i_N}_{k_{1},n_{1},k_{2},n_{2},\cdots,k_{f+1}}=\prod_{i=1}^{f+1} \eta^{-k_{i}}
\Big( \psi_i(z_i)z_i^{k_{i}}+\psi_i(z_i^{-1}) z_i^{-k_{i}} \Big ) \; . 
\end{eqnarray}
It turns out that following constraints have to be fulfilled 
\begin{eqnarray}
&& \frac{\psi_1(z_1)}{\psi_1(z_1^{-1})}=-\frac{F(z_1,\eta,\zeta)}{F(z_1^{-1},\eta,\zeta)}=-z_1^{-2n_{1}} \; ,  \nonumber \\ 
&& \frac{\psi_i(z_i)}{\psi_i(z_i^{-1})}=-z_i^{-2n_{i-1}}=-z_i^{-2n_{i}} \; \mbox{for} \; i=2,\cdots,f \; ,  \nonumber \\ 
&& \frac{\psi_{f+1}(z_{f+1})}{\psi_{f+1}(z_{f+1}^{-1})}=-z_{f+1}^{-2L}\frac{F(z_{f+1}^{-1},\eta^{-1},\xi)}{F(z_{f+1},\eta^{-1},\xi)}=-z_{f+1}^{-2n_{f}}   \nonumber
\end{eqnarray}
and that $z_i$'s have to be obtained from 
\begin{eqnarray}
&& z_1^{2 n_{1}}= \frac{F(z_1^{-1},\eta,\zeta)}{F(z_1,\eta,\zeta)} \;\; , z_{f+1}^{2(L- n_{f})}= \frac{F(z_{f+1}^{-1},\eta^{-1},\xi)}{F(z_{f+1},\eta^{-1},\xi)}  \; ,  \nonumber\\
&& z_{2}^{2(n_{2}-n_{1})}=z_{3}^{2(n_{3}-n_{2})}=\cdots=z_{f}^{2(n_{f}-n_{f-1})}=1 \; .  \nonumber
\end{eqnarray}
These exact results completely determine the eigenvalues and the corresponding left and right eigenvectors of the modified Hamiltonian $\hat{\cal H}(s)$
for any arbitrary $s$. 
\begin{figure}[htp]
\centering
\includegraphics[scale=1.2]{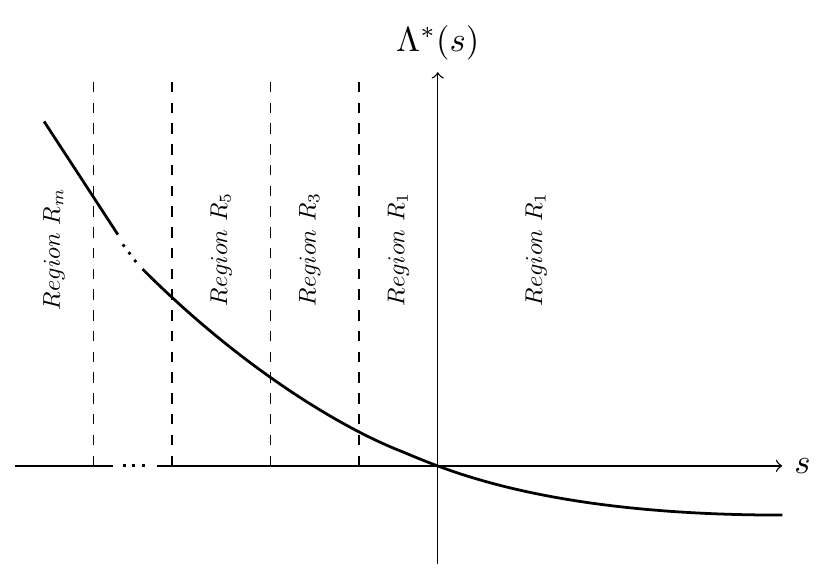}
\caption{The schematic of the maximum eigenvalue of the modified Hamiltonian as a functions of the conjugate field $s$. As can be seen,  
the maximum eigenvalue in each sector (region) is given by a different expression. For more information see inside the text.}
\label{fig2}
\end{figure}
\section{Dynamics of the system conditioned on an atypical value of the activity }
\label{IV}
It is known that the largest eigenvalue of the modified Hamiltonian plays the role of
a dynamical free energy whose discontinuities of its derivatives with respect to $s$
, as a control parameter similar to $\beta=1/kT$ in the traditional statistical physics, 
determine dynamical phase transitions of the system.  
In~FIG.\ref{fig2} we have schematically plotted the largest eigenvalue of the modified Hamiltonian (\ref{new modified hamiltonian}) as a function of $s$. The s-axis 
is divided into different regions because of crossing of the eigenvalues as we have already discussed in~\cite{MTJ14}. Moreover a hierarchy of dynamical phase transitions takes place at the crossing points. 
 The region $R_N$ corresponds to the subspace of $N$ shock fronts
 in which the largest eigenvalue of the modified Hamiltonian is given by the largest element $\{a_{i_N}^{(N)}\}$ in (\ref{ev}) which is 
obtained from (\ref{ev2}). We denote this largest eigenvalue by $\Lambda^{\ast}(s)=a_{i^{{\ast}}_N}^{(N)}$ and the corresponding right and left eigenvectors by
$
| \Lambda^{\ast}(s) \rangle=\bigoplus _{l=1 , l \in odd}^{L+1} |{ \cal{X} }_{i_N^{\ast},l}^{(N)} \rangle
$
and
$
\langle \tilde{\Lambda}^{\ast}(s) |=\bigoplus _{l=1 , l \in odd}^{L+1} \langle{ \tilde{\cal{X}} }_{i_N^{\ast},l}^{(N)} \vert
$
respectively, in which $ |{ \cal{X} }_{i_N^{\ast},l}^{(N)} \rangle$ and $ \langle{ \tilde{\cal{X}} }_{i_N^{\ast},l}^{(N)} \vert$ are 
obtained from (\ref{eigen3}) and (\ref{eigen4}). We can see from (\ref{eigen3}) that the matrix elements of the right eigenvector 
$| \Lambda^{\ast}(s) \rangle$ (the left eigenvector $\langle \tilde{\Lambda}^{\ast}(s) |$) for subspaces with more (less) than $N$ shock fronts 
are zero, so the steady state of the effective Hamiltonian $\hat{\cal{H}}_{eff}(s)$ in $R_N$ is given by
$$
\vert P^{\ast}_{eff}(s) \rangle =\sum_{k_{1}}\sum_{k_{2}}\cdots \sum_{k_{f+1}} 
C^{i^{{\ast}}_N}_{k_{1},n_{1},k_{2},n_{2},\cdots,k_{f+1}} 
 \tilde{C}^{i^{{\ast}}_N}_{k_{1},n_{1},k_{2},n_{2},\cdots,k_{f+1}}
\vert k_{1},n_{1},k_{2},n_{2},\cdots,k_{f+1}\rangle  
$$
in which the coefficients should be obtained from (\ref{coef}).
As can be seen, the steady state of the effective dynamics for a given value of $s$, which
corresponds to the region $R_N$, is written as a linear combination of the product
shock measures with exactly $N$ shock fronts. 

A natural question that might arise is how the particle system organizes itself microscopically
 to produce a rare event with an atypical value of the activity 
and how this microscopic structure can be generated as a typical event 
in the effective dynamics.

Let us consider an arbitrary region, say $R_5$, and study the conditioned and effective dynamics of the system 
conditioned on the values of activity corresponding to this region. From (\ref{eigen2}) and (\ref{eigen3}) one can simply see that the left eigenvector 
$\langle \tilde{\Lambda}^{{\ast}}(s) |$ and the right eigenvector $| \Lambda^{{\ast}}(s) \rangle $ corresponding to the largest eigenvalue of 
modified Hamiltonian have non-zero elements only for the subspaces containing of $l \geq 5$ and $l \leq 5$ shock fronts respectively. 
We conclude the conditioned dynamics 
of the system can be described as follows: the dynamical trajectories of the conditioned process on an atypical value of activity corresponding to
the region $R_5$, start from one of subspaces with $l \geq 5$ shock fronts and end to ones with $l \leq 5$. 
For instance, if a dynamical trajectory starts from the subspace with $9$ shock fronts, 
these shock fronts can merge 
(by annihilation of a random walker and its consecutive obstacle) 
and at the end of this trajectory, the number of shock fronts will be $l \leq 5$. 
\begin{figure}[!t]
\centering
\includegraphics[scale=1]{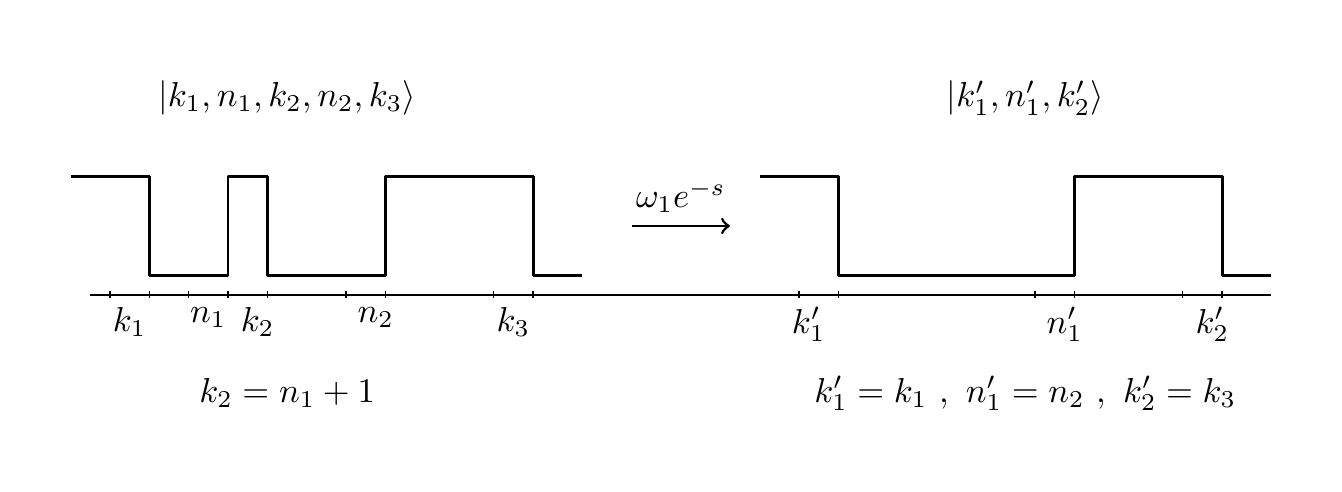}
\caption{A given transition $C \to C'$ in which the configurations $C$ and $C'$ belong to the subspaces with $5$ and $3$ shock fronts, respectively }
\label{fig3}
\end{figure}
In fact, the structure of the non-stochastic evolution generator (\ref{new modified hamiltonian})  allows the shock fronts to merge infinitely.
In contrast, one can see that this is not the case for the generator of 
the effective dynamics $\hat{\cal{H}}_{eff}(s)$. 
There is a limit for merging of shock fronts
in the effective dynamics. Let us show this feature by two simple examples. First, we consider a transition $C \to C'$ according to FIG.\ref{fig3}
in which the configurations $C$ and $C'$ belong to the subspaces with $5$ and $3$ shock fronts, respectively. The matrix element of 
effective Hamiltonian for this transition is given by
$$
\label{element}
\langle C' | \hat{\cal{H}}_{eff}(s) | C \rangle =
\langle C' | \hat{\cal{H}}(s) | C\rangle
\frac
{ \langle \tilde{\Lambda}^{{\ast}}(s) |C' \rangle}
{ \langle \tilde{\Lambda}^{{\ast}}(s) |C \rangle} =
\omega_1 e^{-s}
\frac
{\langle{ \tilde{\cal{X}} }_{i^{{\ast}}_5,3}^{(5)} \vert k'_1,n'_1,k'_2 \rangle}
{\langle{ \tilde{\cal{X}} }_{i^{{\ast}}_5,5}^{(5)} \vert k_1,n_1,k_2,n_2,k_3 \rangle} =0
 \; .
$$

Now, imagine a transition $C'' \to C$ in which the configurations $C''$ and $C$ belong to the subspaces with $7$ and $5$ shock fronts, respectively. 
In this case, we have
$$
\label{element}
\langle C | \hat{\cal{H}}_{eff}(s) | C'' \rangle \neq 0
 \; .
$$
This means that the shock fronts in the effective dynamics can merge provided that the number of them sustain more than $5$ or equal to it; therefore, the effective interaction between shock fronts is long-range. In general, the minimum allowed number of shock fronts in the effective dynamics for $R_N$ is $N$. 
In fact, for generating an amount of activity, there is a specific limit for the number of the shock fronts needed in the effective dynamics. Moreover, the hierarchy of dynamical phase transitions occurs in the system at the crossing points of eigenvalues just as the number of the effective repelling shock fronts on the lattice changes. 

The effective transition rates in the system under studied, which are exactly obtained from (\ref{relation}), (\ref{ev}) and (\ref{eigen2}), have a rather complicated form. However, considering a given value of $s$, which corresponds to the region $R_N$, if the initial state of the system is a linear combination of the product shock measures with $N$ shock fronts, at any later time the state of the system in the effective dynamics 
can be  also expressed in terms of the shock measures with the same number of shock fronts; therefore the effective generator $\hat{\cal{H}}_{eff}(s)$ forbids the number of shock fronts on the lattice to increase or decrease. It means that the microscopic structure of
the system in the effective dynamic evolves in such a way that there seems to be a 
repulsive interaction between all these shock fronts (both random walkers and obstacles) 
and it prevents them from merging. 
This allows us to describe the effective dynamics for a given value of $s$ more easily in terms of the shock fronts on the lattice which repel each other and conserve their number.   
Finally, the steady state probability distribution of effective Hamiltonian of the system 
can also be written in terms of superposition of the product shock measures with
 a certain number of shock fronts depending on the value of the counting field $s$.  

\section{Concluding Remarks }
\label{V}

In this paper, we have studied the activity fluctuations of a variant of the zero-temperature Glauber model in which
 bulk reactions consist of asymmetric death and branching of particles on an one-dimensional lattice with open boundaries.
We have shown that conditioning the model on an atypical value of the activity can be described in terms of the long-range effective interactions between repelling shock fronts defined in (\ref{newbasis}). We have also calculated the effective interaction rates and steady state of the effective dynamics analytically. 

In our previous work~\cite{TJ15a} which was on the activity fluctuations below the typical value in this model, we have shown that the system undergoes 
both continuous and discontinuous dynamical phase transitions. The spatio-temporal patterns that generate such atypical values of activity
over some period of time and different dynamical phases of system in the lower-than-typical activity region
 have been discussed in terms of the configuration of the system at the beginning and the end of each trajectory 
during the observation time. It has been shown that the dynamics of the model conditioned on a lower-than-typical value of activity can be described 
in terms of a single shock front which performs a biased random walk on the lattice. The behavior of the system in the higher-than-typical activity region is different. 
In this region, the system undergoes a hierarchy of dynamical phase transitions and dynamics of the system which is studying in the present work, 
can be described by the evolution of multiple shock fronts on the lattice.

According to the Fig.2, the more shock fronts in the system the more activity is being generated. For instance, in order to produce an atypical value of activity in region $R_N$, 
$N$ shock fronts are involved. More precisely, we have found that the dynamics of the system conditioned on a specific atypical value of activity relates to a specific number of effective shock fronts on the lattice. 
Moreover, a hierarchy of dynamical phase transitions observed in the model, occurs just as the number of these effective shock fronts changes.
We have found that the activity of the system in an arbitrary region, let us say $R_N$, is produced by those trajectories for which the initial state of the system consists of the configurations which contain $N$ or more shock fronts. On the other hand, at the end of the observation time the number of shock fronts might remain unchanged or drop to smaller values than $N$.
On the other hand, These trajectories can be characterized as typical ones for an auxiliary or effective system. Shock fronts in the typical trajectories, in contrast to the trajectories of biased dynamics, can not merge infinitely. 
They can merge provided that their numbers will be not smaller than a specific limit; therefore, the effective interaction between shock fronts are long-range. Moreover, there is a certain number of repelling effective shock fronts on the lattice for producing any amount of activity during the observation time. 

In~\cite {PSS10} the authors have studied the effective dynamics of an asymmetric simple exclusion process on a ring. They have shown that how conditioning the process on carrying an atypical large flux corresponds to avoiding particles of forming clusters and generating an effective repulsive interaction. In our work, conditioning the process on more activities
corresponds to forming a larger number of shock fronts with long-range effective repulsive interaction.

 

\section*{References}
\begin{thebibliography}{99}



\bibitem{rare1} S. Albeverio, V. Jentsch, and H. Kantz, Extreme Events in Nature and Society (Springer, Berlin, 2006).

\bibitem{rare2} S. Auer and D. Frenkel, Nature 409, 6823 (2001); 
                        R. P. Sear, J. Phys.: Cond. Matt. {\bf 19} 033101(2007);
                        E. I. Shakhnovich, A. M. Gutin, Nature {\bf 346} 773 (1990); 
                         E. I. Shakhnovich, Chem. Rev. {\bf 106} 1559 (2006);
                         G. D. Rose, P. J. Fleming, J. R. Banavar, and A. Maritan, 
                         Proc. Natl. Acad. Sci. USA {\bf 103} 16623 (2006).
\bibitem{CT15} R. Chetrite, H. Touchette, Ann. Henri Poincar\'e {\bf16} 2005 (2015).
\bibitem{Ruelle} D. Ruelle, {\it Thermodynamic Formalism} (Addison-Wesley, Reading,  Massachusetts, 1978); J. P. Eckmann, D. Ruelle, Rev. Mod. Phys. 57, 617 (1985).
\bibitem{LAV07} V. Lecomte, C. Appert-Rolland and F. van Wijland, J. Stat. Phys. {\bf127} 51 (2007);J. P. Garrahan, R. L. Jack, V. Lecomte, E. Pitard, K. van Duijvendijk and F. van Wijland, J. Phys. A: Math. Theor. {\bf42} 075007 (2009). 
\bibitem{JS10} R. L. Jack, P. Sollich, Prog. Theor. Phys. Supp. {\bf184} 304 (2010).
\bibitem{PS11} V. Popkov, G. M. Sch\"utz, J. Stat. Phys. {\bf142}  627 (2011).
\bibitem{JS14} R. L. Jack, P. Sollich, J. Phys. A {\bf47} 015003 (2014).
\bibitem{HMS15} O. Hirschberg, D. Mukamel, G. M. Sch\"utz, J. Stat. Mech. P11023 (2015).
\bibitem{JS15} R. L. Jack, P. Sollich, Eur. Phys. J. Special Topics {\bf224} 2351 (2015).
\bibitem{TJ15b} P. Torkaman, F. H. Jafarpour, Phys. Rev. E. {\bf92} 062104 (2015).
\bibitem{TJ15a}  P. Torkaman, F. H. Jafarpour, J. Stat. Mech. P01007 (2015).
\bibitem{S01} G. M. Sch\"utz, \textit{Phase transitions and critical phenomena} vol. 19 3, London: Academic (2001).
\bibitem{HJGC09} L. O. Hedges, R. L. Jack, J. P. Garrahan, and D. Chandler, Science {\bf323} 1309 (2009).    
\bibitem{T09} H. Touchette, Phys. Rep. {\bf 478} 1 (2009).
\bibitem{KhA01} M. Khorrami and A. Aghamohammadi, Phys. Rev. E {\bf 63} 042102 (2001).
\bibitem{KJSh} K. Krebs, F. H. Jafarpour and G. M. Sch\"utz, New Journal of Physics {\bf 5}145.1-145.14 (2003).
\bibitem{HSh07} R. J. Harris, G. M. Sch\"utz, J. Stat. Mech. P07020 (2007).
\bibitem{MTJ14} S. R. Masharian, P. Torkaman, F. H. Jafarpour, Phys. Rev. E. {\bf89} 012133 (2014).
\bibitem{PSS10} V. Popkov, G. M. Sch\"utz, D. Simon, J. Stat. Mech. P10007 (2010).
              
                    

\end{thebibliography}
\end{document}